\documentclass[fleqn,usenatbib]{mnras}

\usepackage{newtxtext,newtxmath}

\usepackage[T1]{fontenc}

\DeclareRobustCommand{\VAN}[3]{#2}
\let\VANthebibliography\thebibliography
\def\thebibliography{\DeclareRobustCommand{\VAN}[3]{##3}\VANthebibliography}

\usepackage{graphicx}	
\usepackage{amsmath}

\usepackage{amssymb}	

\title[Models of high-resolution ALMA strong lenses]{Modelling high-resolution ALMA observations of strongly lensed dusty star forming galaxies detected by \textit{Herschel\thanks{\textit{Herschel} is an ESA space observatory with science instruments provided by European-led Principal Investigator consortia and with important participation from NASA}}}

\author[J. Maresca et al.]{
Jacob Maresca,$^{1}$\thanks{E-mail: jacob.maresca@nottingham.ac.uk}
Simon Dye,$^{1}$
Aristeidis Amvrosiadis,$^{2}$
George Bendo,$^{3}$
Asantha Cooray,$^{4}$\newauthor
Gianfranco De Zotti,$^{5}$
Loretta Dunne,$^{6}$
Stephen Eales,$^{6}$
Cristina Furlanetto,$^{7}$
Joaquin Gonz{\'a}lez-Nuevo,$^{8,9}$\newauthor
Michael Greener,$^{1}$
Robert Ivison,$^{10}$
Andrea Lapi,$^{11}$
Mattia Negrello,$^{6}$
Dominik Riechers,$^{12, 13}$\newauthor
Stephen Serjeant,$^{14}$
M\^onica Tergolina,$^{7}$
Julie Wardlow$^{15}$\\
$^{1}$School of Physics \& Astronomy, University of Nottingham, University Park, Nottingham, NG7 2RD, UK\\
$^{2}$Centre for Extragalactic Astronomy, Durham University, Department of Physics, South Road, Durham DH1 3LE, UK\\
$^{3}$UK ALMA Regional Centre Node, Jodrell Bank Centre for Astrophysics, Department of Physics and Astronomy, University of Manchester, Oxford Road, \\ \ Manchester, M13 9PL, UK\\
$^{4}$Department of Physics \& Astronomy, University of California, Irvine CA 92697, USA\\
$^{5}$INAF--Osservatorio Astronomico di Padova, Vicolo dell'Osservatorio 5, I-35122, Padova, Italy\\
$^{6}$School of Physics and Astronomy, Cardiff University, Queens Buildings, The Parade, Cardiff, CF24 3AA, UK\\
$^{7}$Instituto de F{\'i}sica, Universidade Federal do Rio Grande do Sul, Av. Bento Gon\c{c}alves, 9500, 91501-970, Porto Alegre, Brazil\\
$^{8}$Departamento de Fisica, Universidad de Oviedo, C. Federico Garcia Lorca 18, 33007 Oviedo, Spain\\
$^{9}$Instituto Universitario de Ciencias y Tecnologías Espaciales de Asturias (ICTEA), C. Independencia 13, 33004 Oviedo, Spain\\
$^{10}$European Southern Observatory, Karl-Schwarzschild-Strasse 2, D-85748 Garching bei M\"{u}nchen, Germany\\
$^{11}$SISSA, Via Bonomea 265, 34136 Trieste, Italy\\
$^{12}$Department of Astronomy, Cornell University, Space Sciences Building, Ithaca, NY 14853, USA\\
$^{13}$Max-Planck-Institut f\"{u}r Astronomie, K\"{o}nigstuhl 17, D-69117 Heidelberg, Germany\\
$^{14}$Department of Physical Sciences, The Open University, Walton Hall, Milton Keynes, MK7 6AA, UK\\
$^{15}$Department of Physics, Lancaster University, Lancaster, LA1 4YB, UK
}

\date{Accepted XXX. Received YYY; in original form ZZZ}

\pubyear{2021}

\begin{document}
\label{firstpage}
\pagerange{\pageref{firstpage}--\pageref{lastpage}}
\maketitle

\begin{abstract}
We present modelling of $\sim 0.1$ arcsec resolution Atacama Large Millimetre/sub-millimeter Array imaging of seven strong gravitationally lensed galaxies detected by the Herschel Space Observatory. Four of these systems are galaxy-galaxy scale strong lenses, with the remaining three being group-scale lenses. Through careful modelling of visibilities, we infer the mass profiles of the lensing galaxies and by determining the magnification factors, we investigate the intrinsic properties and morphologies of the lensed sub-millimetre sources. We find that these sub-millimetre sources all have ratios of star formation rate to dust mass that is consistent with or in excess of the mean ratio for high-redshift sub-millimetre galaxies and low redshift ultra-luminous infrared galaxies. The contribution to the infrared luminosity from possible AGN is not quantified and so could be biasing our star formation rates to higher values. The majority of our lens models have mass density slopes close to isothermal, but some systems show significant differences.
\end{abstract}




\section{Introduction}

Sub-millimetre (sub-mm) galaxies (SMGs) \citep[see][for a review]{casey} play host to some of the most intense star formation rates in the Universe. They are observed to be abundant at redshifts $z > 1$ and to contribute approximately 20 per cent of the cosmic star formation rate density up to a redshift of $z \sim 4$ \citep{swinbank, lapi_sfr}. The rest-frame UV and optical radiation associated with young stars is highly obscured by the dust content of sub-mm galaxies \citep{dudzevic}. Dust grains absorb this radiation, causing them to be heated. This energy is reprocessed and emitted as thermal continuum emission that we can observe in the sub-mm and mm regimes. Sub-mm galaxies inhabit a key role in our picture of galaxy evolution; massive SMGs at high-redshift evolve through the population of quiescent galaxies observed at lower redshift \citep{simpson, toft}, and through the mechanism of gas-poor minor mergers, helps to explain the build up of massive elliptical galaxies observed in the present day Universe \citep{oogi, guo, lapi_evolution}. In the local Universe, Ultra Luminous Infrared Galaxies (ULIRGs), are often seen as analogues to the high-redshift sub-mm galaxies due to their strongly dust-obscured UV luminosity and high infrared luminosity, and whilst they are considerably less abundant, they have comparable bolometric luminosities \citep{alaghband, rowlands}. The study of high-redshift sub-mm galaxies has benefited greatly from the advent of large interferometer arrays such as the Atacalma Large Millimeter/Sub-millimeter Array (ALMA), allowing observations to reach resolutions of $<$ 0.1 arcsec, probing physical scales that were previously unreachable.

Strong gravitational lensing provides a further boost in spatial resolution due to the magnification of the background source, which is typically within the range of 5-10 for SMGs \citep{spilker}. In addition to this, there exists a strong lensing bias in the sub-mm regime, making it possible to find lensed sources in wide surveys with a simple cut in flux density above 100 $\mathrm{mJy}$ at 500 $\mu$m \citep{blain, negrello_2007, negrello_2010, perrotta}. Using this technique, follow-up ALMA observations of strongly lensed sub-mm galaxies detected in wide area surveys, such as the Herschel Astrophysical Terahertz Large Area Survey \citep[H-ATLAS;][]{hatlas}, the Herschel Multi-tiered Extragalactic Survey \citep[HerMES;][]{hermes} and the Herschel String 82 Survey \citep[HerS;][]{Viero_2014}, carried out using the Herschel Space Observatory \citep{hso}, as well as the sub-mm surveys conducted by the Planck satellite \citep{planck_canameras} and the millimetre surveys of the South Pole Telescope \citep{SPT} have contributed to a rapidly increasing understanding of galaxy formation at its early stages \citep{dye_2018, Sun_2021, harrington, gems3, gems4}. Many earlier observations were focused on extremely luminous sources, but with the increased sensitivity in instruments such as ALMA, it is becoming possible to investigate more typical main-sequence star-forming galaxies.

High-resolution sub-mm follow-up observations allow for precision lensing analyses, greatly benefiting our characterisation of source properties sensitive to the lens model, such as the intrinsic luminosity, star formation rate and gas and dust masses. Studies of the molecular gas and dust in lensed sub-mm galaxies  allow us to test models of star formation in the early Universe \citep{cava, dessauges}. Recent studies have begun to reveal the compact nature of dust in sub-mm galaxies \citep{puglisi, tadaki} and the disparity in size of these regions when compared with local ULIRGs. Similarities in size, number density and clustering properties with compact quiescent galaxies is suggestive of an evolutionary connection \citep{An, dudzevic}, and thus understanding of quenching in sub-mm galaxies is important to understand how they may become red and dead. The advent of observatories such as ALMA have revolutionised our understanding of high-redshift lensed sources when compared to what could be achieved with optical imaging. A classic example of this is the lensed system SDP.81, first discovered within the H-ATLAS sample and then observed with ALMA \citep{dye_2015, 2015_rybak, rybak_2015, negrello_2010}.

As our sample size of known strongly lensed sub-mm galaxies increases, so does our range of redshifts at which they have been observed \citep{wang, riechers} thanks to a very negative K-correction such that high-redshift galaxies have approximately constant brightness in the sub-mm regime between redshifts $z \sim 1 - 8$. Higher redshift sources tend to be lensed by higher redshift lenses, due to the scaling of the lensing cross-section with redshift \cite{cross_section}. Increasing the redshift range then allows us to study the mass profiles of lenses at earlier epochs, when galaxy evolution is more rapid and less well understood \citep{dye_2014, negrello_2014}.

Reconstruction of the background lensed source from interferometer observations can be achieved with two distinct approaches. There are methods that model the visibilities directly in the uv-plane \citep{bussmann_2012, bussmann_2013, dye_2018}, and those that model the cleaned image plane data \citep{dye_2015, Inoue, yang}. The benefit of working in the image plane is that the task is significantly less computationally expensive, however, due to the incomplete coverage of the uv-plane, spatially correlated noise is introduced which can in principle bias the inferred lens model. Working directly with the visibility data avoids this problem at the cost of longer modelling times.

In this work, we have carried out lens modelling in the uv-plane of seven ALMA observations, four of which are galaxy-galaxy scale, and the remaining three are group-scale lenses. These systems were originally detected by Herschel within H-ATLAS and the extension to the HerMES field, HerMES Large Mode Survey (HeLMS) \citep{asboth, Nayyeri_2016}. We have investigated the intrinsic source properties, such as luminosity and SFR, by determining the magnification factors. Additionally, we investigate the morphologies of the reconstructed sources.

The layout of this paper is as follows: Section \ref{sec:data} describes the ALMA observations and other sources that were drawn upon for this work. Section \ref{sec:method} details the methodology of our lens modelling and Section \ref{sec:results} presents the results of our work. In Section \ref{sec:discussion} we compare our results to other similar studies. Finally, in Section \ref{sec:conclusions} we summarise our findings and discuss their interpretation. Throughout this paper we assume a flat $\Lambda\mathrm{CDM}$ cosmology using the 2015 Planck results \citep{planck_2015}, with Hubble parameter $h=0.677$ and matter density parameter $\Omega_m = 0.307$.

\section{DATA}
\label{sec:data}

The seven ALMA observations modelled in this work are from the ALMA programme 2013.1.00358.S (PI: Stephen Eales) and are described in detail within \cite{aris}. The observation targets for the original ALMA programme were selected from the H-ATLAS and HeLMS surveys for having the brightest 500 $\mu\mathrm{m}$ flux densities and with spectroscopic redshifts > 1. Of the 16 sources observed during ALMA cycle 2, 14 display obvious lensing features. Six of those sources are modelled in the work by \cite{dye_2018}, one of them we leave for future work, and the remaining seven are modelled here. Of these seven sources, one was identified in H-ATLAS whilst the remaning six are from HeLMS. 

The spectral setup employed by ALMA was identical for each of the lensing systems observed. The band 7 continuum observations were comprised of four spectral windows, each with a width of 1875 MHz and centred on the frequencies 336.5 GHz, 338.5 GHz, 348.5 GHz and 350.5 GHz. The central frequency of 343.404 GHz corresponds to a wavelength of 873 $\mu$m. Each spectral window is comprised of 128 frequency channels, resulting in a spectral resolution of 15.6 MHz. The ALMA configuration utilised forty two 12m antennae with an on-source integration time of approximately 125 s. Upon combining all four spectral windows, this achieves an angular resolution of 0.12 arcsec and RMS values of approximately 230 $\mu$Jy/beam and 130 $\mu$Jy/beam for the H-ATLAS and HeLMS sources, respectively. The synthesised beam shape for HeLMS J005159.4+062240 is the least elliptical of our sample, with a ratio of major to minor axis of $\sim 1.2$. For the observation of HeLMS J235331.9+031718, the ratio of major to minor axis of the synthesised beam is the most elliptical of our sample at $\sim 1.7$, with the remaining observations having ratios of $\sim 1.5$. The maximum recoverable scales for our observations are between 1.32 arcsec and 1.46 arcsec. 

In this work, we have used the visibility data provided by the ALMA science archive, and re-calibrated them using \textsc{Common Astronomy Software Applications} version (CASA) 4.3.1 \citep{mcmullin} and the scripts provided by the archive. Baselines flagged as bad by the ALMA data reduction pipeline were excluded from the analysis. The CASA task tclean was used to create images in order to measure the flux of the sources at 873 $\mu$m. The images were constructed using a natural weighting scheme and were primary beam corrected. To well sample the minor axis of the primary beam, an image pixel scale of 0.02 arcsec and 0.03 arcsec was used for the H-ATLAS and HeLMS sources respectively. For the creation of these images, we used CASA version 6.1.2.7.

For the calculation of intrinsic source properties, photometry from our ALMA data was used in combination with a number of other datasets. Sub-mm photometry, obtained by the Herschel Space Observatory, making use of both the Spectral and Photometric Imaging Receiver \citep[SPIRE;][]{griffin} at wavelengths of 250 $\mu$m, 350 $\mu$m and 500 $\mu$m and the Photoconductor Array Camera and Spectrometer \citep[PACS;][]{poglitsch} at wavelengths of 100 $\mu$m and 160 $\mu$m. SPIRE and PACS photometry for the H-ATLAS and HeLMS sources was taken from \cite{zhang} and \cite{Nayyeri_2016}. Where possible, we have also used 850 $\mu$m Submillimeter Common User Bolometer Array 2 (SCUBA-2) fluxes (1 arcsec resolution) taken from \cite{bakx} and 880 $\mu$m photometry taken from the Submillimeter Array (SMA) (0.6 arcsec resolution) as described in \cite{bussmann_2013}. Finally, the ALMA band 4 (1940 $\mu$m, 0.3 arcsec resolution) flux for H-ATLAS J083051.0+01322 was taken from \cite{yang}.

The available lens and source redshifts for the seven systems modelled in this paper can be found in Table \ref{tab:redshifts}. The observed photometry can be found in Table \ref{tab:photometry}.

\begin{table}
	\centering
	\caption{The list of the seven lensing systems modelled in this work, along with their lens galaxy redshifts, $z_{l}$, and their background source redshifts $z_{s}$. Where Appropriate, the redshift of both lensing galaxies have been provided, distinguished by the numbered subscript. The references from which the lens and source redshifts were obtained are as follows: $^{\mathrm a}$\protect\cite{bussmann_2013}, $^{\mathrm b}$\protect\cite{negrello_2016}, $^{\mathrm{c}}$\protect\cite{Nayyeri_2016}, and $^{\mathrm{d}}$\protect\cite{z_helms18}. The range of redshifts given for HeLMS J235331.9+031718 is based on the range of source redshifts in this paper, since there is no available redshift measurement for this source. A dash indicates missing redshift information for a lens whilst 'N/A' is used to indicate a second mass profile was not used in our modelling procedure.}
	\label{tab:redshifts}
	\begin{tabular}{lccr} 
		\hline
		ID & $z_{l,1}$ & $z_{l,2}$ & $z_s$ \\
		\hline
		H-ATLAS J083051.0+013225 & 0.626$^\mathrm{b}$ & 1.002$^\mathrm{b}$ & 3.634$^\mathrm{a}$ \\
		HeLMS J005159.4+062240 & 0.602$^{\mathrm{d}}$ & 0.599$^{\mathrm{d}}$ & 2.392$^{\mathrm{c}}$ \\
		HeLMS J234051.5-041938 & - & N/A & 3.500$^{\mathrm{c}}$ \\
		HeLMS J232439.5-043935 & - & - & 2.473$^{\mathrm{c}}$ \\
		HeLMS J233255.4-031134 & 0.426$^{\mathrm{c}}$ & N/A & 2.690$^{\mathrm{c}}$ \\
		HeLMS J233255.6-053426 & 0.976$^{\mathrm{c}}$ & N/A & 2.402$^{\mathrm{c}}$ \\
		HeLMS J235331.9+031718 & 0.821$^{\mathrm{c}}$ & N/A & 2.0 -- 3.7 \\
		\hline
	\end{tabular}
\end{table}

\begin{table*}
	\centering
	\caption{Observed (i.e lensed) source flux densities in mJy. The passband central wavelength in $\mu$m is indicated by the subscripts. For both the H-ATLAS and HeLMS sources, the flux densities $f_{100}$ to $f_{500}$ are taken from \protect\cite{zhang} and \protect\cite{Nayyeri_2016}. The flux densities $f_{850}$, $f_{873}$, $f_{880}$ and $f_{1940}$ are taken from \protect\cite{bakx}, this work, \protect\cite{bussmann_2013} and \protect\cite{yang}, respectively.}
	\label{tab:photometry}
	\begin{tabular}{lccccccccr} 
		\hline
		ID & $f_{100}$ & $f_{160}$ & $f_{250}$ & $f_{350}$ & $f_{500}$ & $f_{850}$ & $f_{873}$ & $f_{880}$ & $f_{1940}$ \\
		\hline
		H-ATLAS J083051.0+013225 & 53 $\pm$ 3 & 198 $\pm$ 10 & 260 $\pm$ 7 & 321 $\pm$ 8 & 269 $\pm$ 9 & 121 $\pm$ 9 & 92 $\pm$ 9 & 86 $\pm$ 4 & 8.8 $\pm$ 0.5 \\
		HeLMS J005159.4+062240 & 31 $\pm$ 3 & 91 $\pm$ 15 & 166 $\pm$ 6 & 195 $\pm$ 6 & 135 $\pm$ 7 & - & 41 $\pm$ 4 & - & - \\
		HeLMS J234051.5-041938 & 7 $\pm$ 3 & 68 $\pm$ 7 & 151 $\pm$ 6 & 209 $\pm$ 6 & 205 $\pm$ 8 & - & 94 $\pm$ 9 & - & - \\
		HeLMS J232439.5-043935 & 33 $\pm$ 4 & 129 $\pm$ 7 & 214 $\pm$ 7 & 218 $\pm$ 7 & 172 $\pm$ 9 & - & 36 $\pm$ 4 & - & - \\
		HeLMS J233255.4-031134 & 25 $\pm$ 4 & 146 $\pm$ 14 & 271 $\pm$ 6 & 336 $\pm$ 6 & 263 $\pm$ 8 & - & 75 $\pm$ 8 & - & - \\
		HeLMS J233255.6-053426 & 14 $\pm$ 3 & 44 $\pm$ 8 & 148 $\pm$ 6 & 187 $\pm$ 6 & 147 $\pm$ 9 & - & 48 $\pm$ 5 & - & - \\
		HeLMS J235331.9+031718 & - & - & 102 $\pm$ 6 & 123 $\pm$ 7 & 111 $\pm$ 7 & - & 26 $\pm$ 3 & - & - \\
		\hline
	\end{tabular}
\end{table*}

\section{Methodology}
\label{sec:method}

\subsection{The semi-linear inversion method in the uv-plane}

The standard image plane approach of the semi-linear inversion (SLI) method makes use of a pixelised source plane. For a given lens model, the image of each pixel is formed and the linear superposition of these images that best fits the data determines the source surface brightness distribution. Analogously to the image plane version, when working with interferometer visibility data, a model set of visibilities is formed for an image of each source pixel. The linear combination of these model visibilities determines the source surface brightness distribution for a particular lens model.

We used the source inversion method implemented within \texttt{PyAutoLens} \citep{Nightingale2021} which is based on the operator approach described within \cite{powell}. An interferometer visibility dataset $\boldsymbol d$ is comprised of samples of complex visibilities. The surface brightness in the source plane is given by the vector $\boldsymbol s$, with each element corresponding to the surface brightness of a source plane pixel. The parameterised projected surface mass density of the lens model is given by the vector $\boldsymbol{\eta}$ and the mapping of the source light $\boldsymbol s$ to the image plane is described by the operator $\bf L(\boldsymbol{\eta})$. The sky brightness is therefore simply $\bf{L}(\boldsymbol \eta)\bf{s}$. The response of an interferometer is encoded into the operator $\bf D$, which performs the Fourier transforms to convert the pixelised sky brightness distribution into a set of complex visibilities. The observed data $\boldsymbol d$, can therefore be described by the combination of these effects 
\begin{equation}
    \boldsymbol d = \bf D \bf{L}(\boldsymbol \eta)\bf{s} + \bf{n}.
\end{equation}
Assuming uncorrelated noise $\boldsymbol n$ in the visibility data, the diagonal matrix $\bf{C}^{-1}$ represents the covariance. Using a similar method to \cite{dye_2018} to determine the 1$\sigma$ uncertainties on the visibilities, we used the \texttt{CASA} task \texttt{statwt} to empirically measure the visibility scatter computed over all baselines.

Combining this with the set of model visibilities $\bf D \bf L(\boldsymbol \eta) \boldsymbol s$ allows us to write the $\chi^2$ statistic as 
\begin{equation}
    \chi^2 = (\bf D \bf L \boldsymbol s - d)^T \bf{C}^{-1} (\bf D \bf L \boldsymbol s - d).
\end{equation}
With the addition of a prior on the source denoted by the operator $\bf R$ and with the regularisation strength given by $\lambda_s$, \cite{powell} show that we can write the regularised least-squares equation as
\begin{equation}
    \left[(\bf{D} \bf{L})^T \bf{C}^{-1}\bf{D}\bf{L} + \lambda_{s} \bf{R}^T \bf{R}\right]\boldsymbol{s}_{\mathrm{MP}} = (\bf{D} \bf{L})^T\bf{C}^{-1}\boldsymbol{d}
    \label{eq:maps}
\end{equation}
where the maximum a posteriori source inversion matrix, i.e the solution matrix for $\boldsymbol{s}_{\mathrm{MP}}$ is given by the quantity in square brackets in equation \ref{eq:maps}. This linear system of equations is in principle straightforward to solve but becomes extremely memory intensive for large numbers of visibilities and/or large numbers of source plane pixels. For this reason, a Direct Fourier Transform (DFT) is replaced by a non-uniform Fast Fourier Transform (NUFFT), constructed out of a series of operators. In \texttt{PyAutoLens}, the Fourier transforms are performed using the NUFFT algorithm \texttt{PyNUFFT} \citep{pynufft}, which greatly increases computational efficiency over the DFT method. This substitution results in a modified version of equation \ref{eq:maps}, where the solution for $\boldsymbol{s}_{\mathrm{MP}}$ is given by a series of operators evaluated by use of an iterative linear solver. The linear algebra package \texttt{PyLops} \citep{pylops}, is used to achieve this \cite[see][for more details on this methodology and the specific implementation used here]{powell, Nightingale2021}.

To find the optimal lens model parameters with \textsc{PyAutoLens}, we used the nested sampling algorithm \textsc{dynesty} \citep{dynesty} to maximise the bayesian evidence as derived within \cite{suyu}. We adopted a gradient regularisation scheme, analogous to that which is described in \cite{warren_dye}, with a constant weight for the source plane due to it's simplicty and to not add more computational costs to an already expensive procedure. We first reconstructed the background source using a source plane pixelisation adapted to the lens magnification. This magnification-based fit was then used to initialise a new search of parameter space with a source plane that adapted to the brightness of the reconstructed source \citep{Nightingale2015, Nightingale2018}. The mass model parameters, along with the source plane parameters (regularisation, number of source plane pixels), were fully optimised throughout the lens modelling procedure.

\subsection{Lens Model}

We have used the elliptical power-law density profile, which is a generalised form of the singular isothermal ellipsoid that is commonly used to fit strong lens profiles \citep{keeton2001catalog}. When it improves the Bayesian evidence, an external shear component is included to compensate for the influence of line of sight galaxies that may be outside our field of view. Where necessary, two elliptical power-law profiles have been used to model the group-scale lenses present in the sample. We find that in all cases this is sufficient to provide acceptable fits to the data and through the use of optical/near-IR imaging, we see no indication of more than two lenses being required to produce an accurate model. The surface mass density, $\kappa$, of this profile is given by

\begin{equation}
\centering
    \kappa(R) = \frac{3 - \alpha}{1 + q} \left(\frac{\theta_E}{R}\right)^{\alpha - 1},
	\label{eq:powerlaw}
\end{equation}

where $\theta_E$ is the model Einstein radius in arc seconds, $\alpha$ the power-law index and $R$ the elliptical radius defined as $R = \sqrt{x^2 + y^2 / q^2}$, where $q$ is the axis ratio \citep{suyu_lens_model}. The lens orientation $\phi$, is measured as the angle counter-clockwise from the east axis to the semi-major axis of the elliptical lens profile. The centre of the lens profile is given by the coordinates in the image-plane $(x_c, y_c)$. The external shear field is parameterised by a shear strength $\gamma$ and shear direction $\phi_{\gamma}$, measured counter-clockwise from east. The shear direction is defined to be perpendicular to the direction of the resulting image stretch. For single lens systems there are six lens model parameters (eight when including external shear) and 12 lens model parameters (14 when including external shear) in the case of group-scale lenses.

\section{Results}
\label{sec:results}

\begin{table*}
	\centering
	\caption{Lens model parameters. The columns are the Einstein radius, $\theta_E$, coordinates ($x_c, y_c$) of the centroid of the mass profile with respect to the phase-tracking centre of the observations ($x_c$ and $y_c$ correspond to west and north respectively), the lens profile orientation measured as the angle counter clockwise from East to the semi major axis, $\phi$, the lens profile axis ratio, $q$, the density slope of the power-law, $\alpha$, the magnitude of the external shear field, $\gamma$, and the external shear field direction, $\phi_{\gamma}$, measured counter clockwise from east}
	\label{tab:lens_models}
	\begin{tabular}{lccccccr} 
		\hline
		ID & $\theta_E$ (arcsec) & $(x_c, y_c)$ (arcsec) & $\phi$ (deg) & $q$ & $\alpha$ & $\gamma$ & $\phi_{\gamma}$ (deg) \\
		\hline
		H-ATLAS J083051.0+013225 (lens 1) & 0.43 $\pm$ 0.01 & (-1.87 $\pm$ 0.01, -0.78 $\pm$ 0.01) & 7 $\pm$ 1 & 0.51 $\pm$ 0.01 & 2.18 $\pm$ 0.02 & - & - \\
		H-ATLAS J083051.0+013225 (lens 2) & 0.54 $\pm$ 0.01 & (-1.51 $\pm$ 0.01, -0.08 $\pm$ 0.01) & 9 $\pm$ 2 & 0.90 $\pm$ 0.01  & 2.03 $\pm$ 0.08 & - & - \\
		HeLMS J005159.4+062240 (lens 1) & 3.80 $\pm$ 0.02 & (-4.67 $\pm$ 0.01, 4.02 $\pm$ 0.01) & 156 $\pm$ 2 & 0.37 $\pm$ 0.05 & 2.05 $\pm$ 0.03 & 0.17 $\pm$ 0.02 & 102 $\pm$ 3 \\
		HeLMS J005159.4+062240 (lens 2) & 1.46 $\pm$ 0.02 & (-0.63 $\pm$ 0.01, 2.18 $\pm$ 0.01) & 109 $\pm$ 5 & 0.66 $\pm$ 0.06 & 1.93 $\pm$ 0.02 & - & - \\
		HeLMS J234051.5-041938 & 0.75 $\pm$ 0.01 & (-0.78 $\pm$ 0.01, -1.88 $\pm$ 0.01) & 97 $\pm$ 1 & 0.27 $\pm$ 0.01 & 2.37 $\pm$ 0.01 & 0.17 $\pm$ 0.01 & 103 $\pm 1$ \\
		HeLMS J232439.5-043935 (lens 1) & 0.54 $\pm$ 0.01 & (1.91 $\pm$ 0.02, 1.09 $\pm$ 0.02) & 23 $\pm$ 1 & 0.79 $\pm$ 0.02 & 2.08 $\pm$ 0.06 & 0.13 $\pm$ 0.01 & 165 $\pm$ 2 \\
		HeLMS J232439.5-043935 (lens 2) & 0.40 $\pm$ 0.01 & (0.88 $\pm$ 0.01, 1.24 $\pm$ 0.01) & 84 $\pm$ 2 & 0.27 $\pm$ 0.02 & 1.84 $\pm$ 0.03 & - & - \\
		HeLMS J233255.4-031134 & 1.01 $\pm$ 0.01 & (-0.06 $\pm$ 0.01, -1.52 $\pm$ 0.02) & 118 $\pm$ 2 & 0.44 $\pm$ 0.02 & 1.89 $\pm$ 0.01 & 0.18 $\pm$ 0.01 & 112 $\pm$ 1 \\
		HeLMS J233255.6-053426 & 1.05 $\pm$ 0.01 & (2.43 $\pm$ 0.01, 0.18 $\pm$ 0.01) & 150 $\pm$ 1 & 0.56 $\pm$ 0.01 & 1.99 $\pm$ 0.03 & 0.12 $\pm$ 0.01 & 48 $\pm$ 1 \\
		HeLMS J235331.9+031718 & 0.21 $\pm$ 0.01 & (1.81 $\pm$ 0.01, 1.32 $\pm$ 0.01) & 151 $\pm$ 4 & 0.49 $\pm$ 0.05 & 1.64 $\pm$ 0.04 & 0.19 $\pm$ 0.02 & 170 $\pm$ 5 \\
		\hline
	\end{tabular}
\end{table*}

\begin{figure*}
	\includegraphics[width=0.99\textwidth]{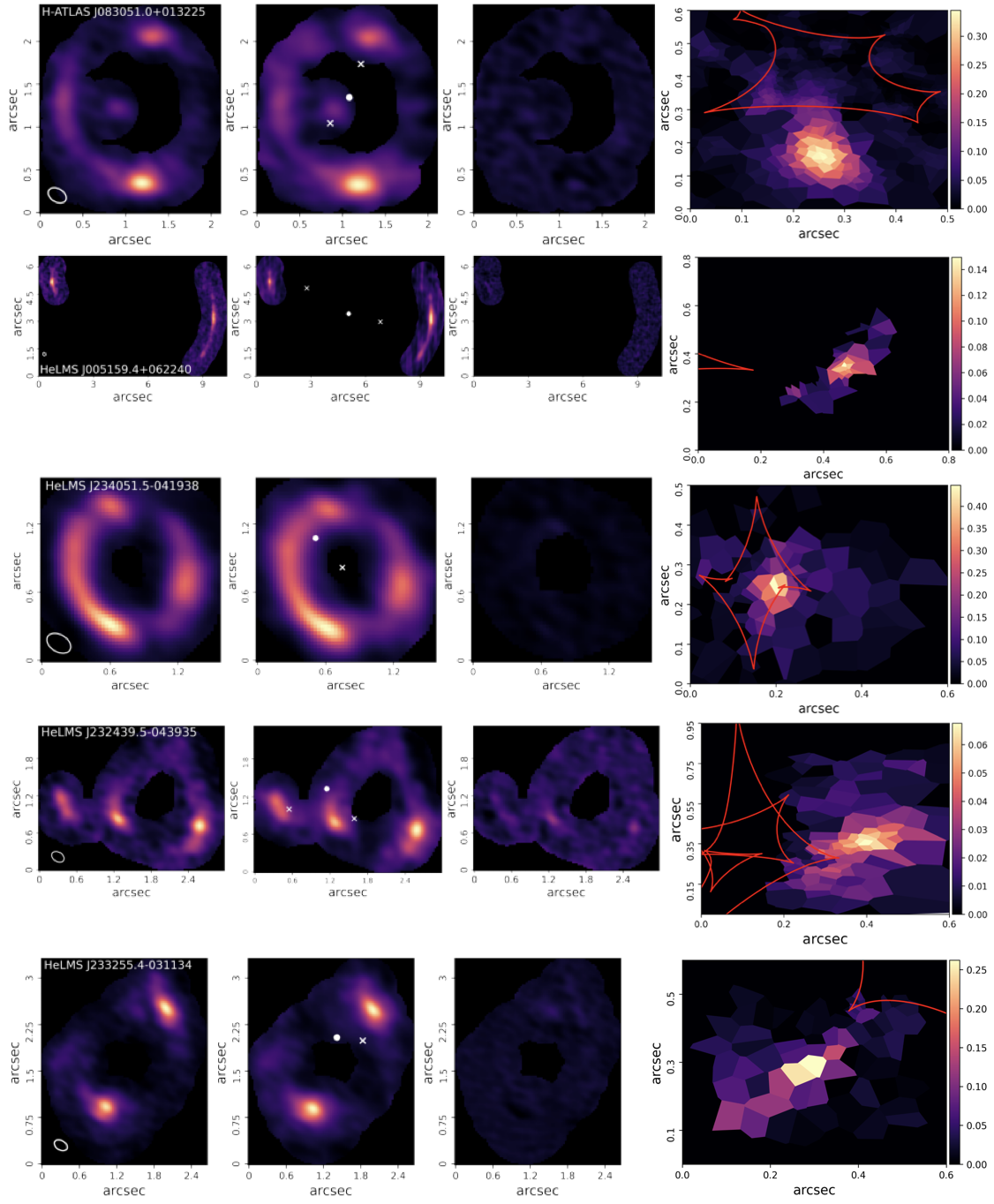}
    \caption{Lens reconstructions for each system. The left column shows the cleaned ALMA image. The middle left column shows the lensed image of our reconstructed source, i.e the model image with the lens centroid(s) and centre of the source plane indicated by white crosses and white circles respectively. The middle right column shows the residual image for our model, i.e the image of the observed visibilities minus the image of the model visibilities. The right column shows the reconstructed source. The colour bar indicates the 873$\mu$m surface brightness in Jy/arcsec$^2$ for each of the panels related to a particular source and all residuals are $< 3 \sigma$. The caustics are shown in red. North is up and East is left.}
    \label{fig:model_plots}
\end{figure*}

\renewcommand{\thefigure}{\arabic{figure} (Continued)}
\addtocounter{figure}{-1}

\begin{figure*}

	\includegraphics[width=\textwidth]{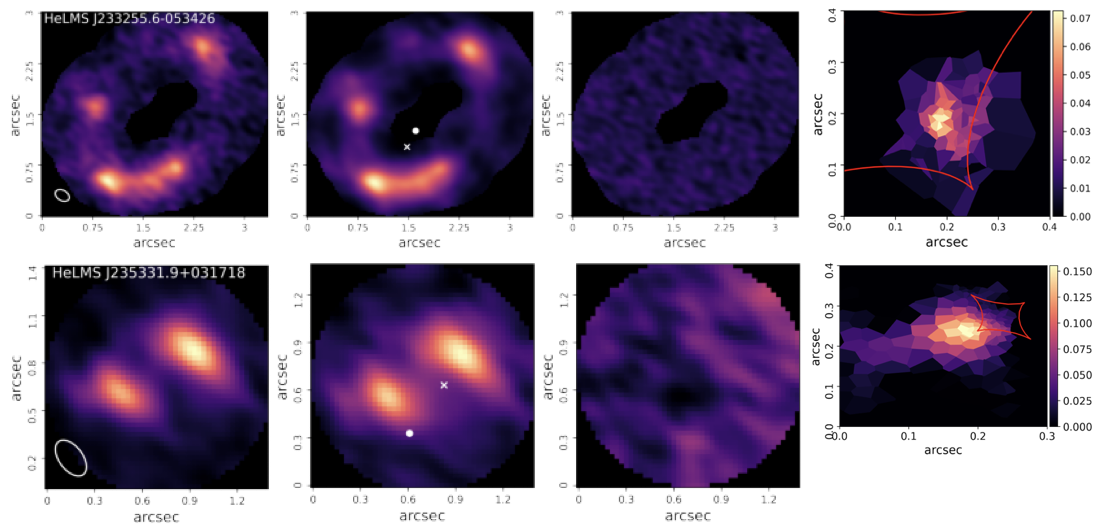}
    \caption{}
\end{figure*}

\renewcommand{\thefigure}{\arabic{figure}}

Fig \ref{fig:model_plots} shows the model image, residual image and source reconstruction for each of the seven lenses modelled in this work. The lens model parameters from our fitting procedure are given in Table \ref{tab:lens_models}.

\begin{figure*}
	\includegraphics[width=\textwidth]{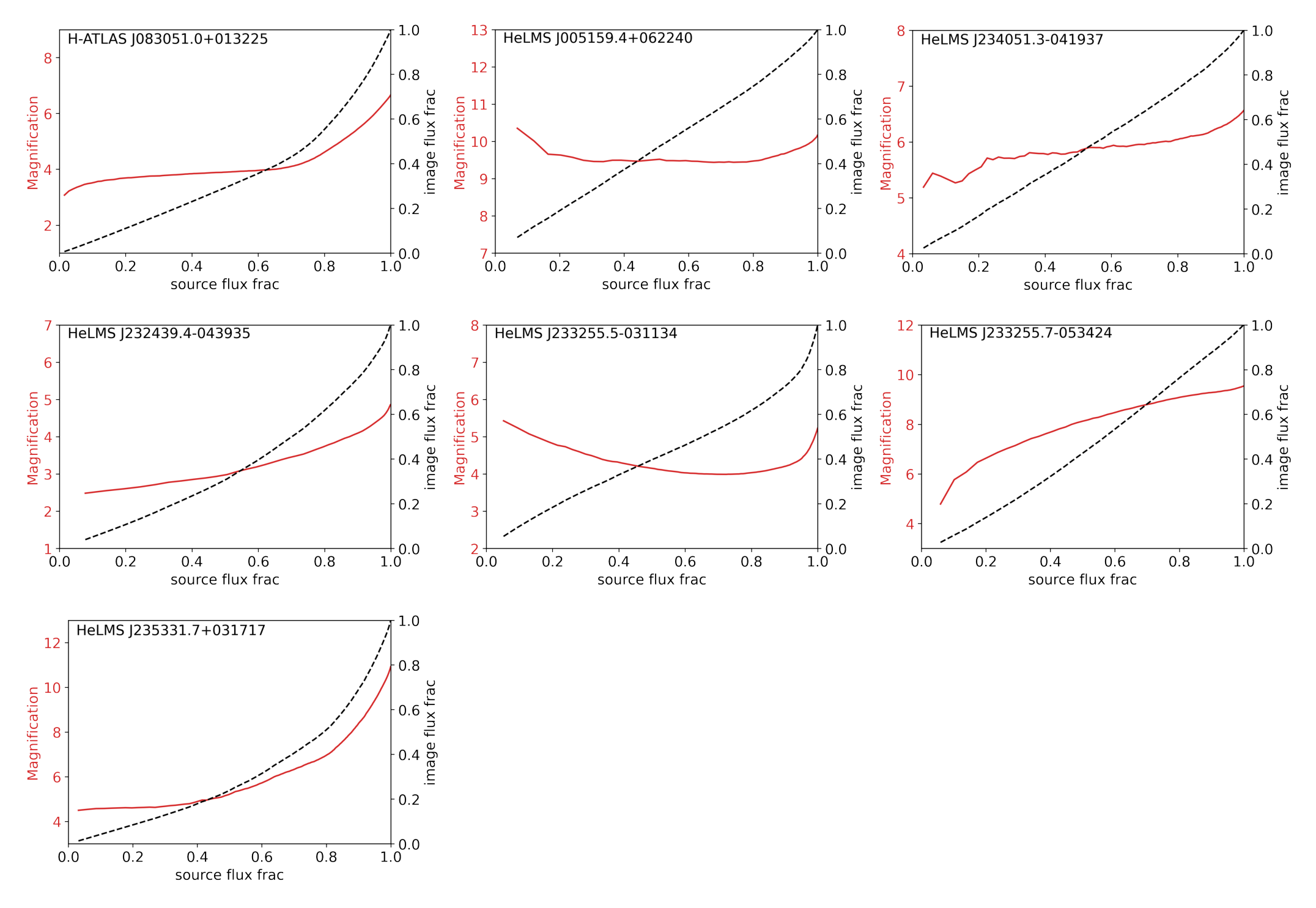}
    \caption{The magnification profile for each of the reconstructed background sources. Each plot shows how the total magnification (solid red line) and image flux density fraction (dashed black line) varies as a function of the source flux density fraction above a surface brightness threshold (more details given in the main text). The magnification profiles represent the average of 100 realisations of the source plane pixelisation for the best-fit lens model. The plot demonstrates how the inferred magnification can vary with different inteferometer configurations.}
    \label{fig:mag_profile_plots}
\end{figure*}

Different interferometer configurations are able to probe distinct scales and reach varying surface brightness limits. Fig \ref{fig:mag_profile_plots} shows how for each system the inferred magnification is sensitive to this effect. Working down a list of source plane pixels, above a surface brightness threshold, ranked by flux density (i.e the product of their reconstructed surface brightness and area), the reconstructed image and average magnification was computed. The source flux fraction refers to the fraction of the total flux contained by a subset of these ranked pixels, e.g five pixels with the greatest flux might contain one third of the total flux of all the pixels in the source plane above a surface brightness threshold. Likewise, the image flux fraction refers to the fraction of the total flux in the image plane that these subsets of source plane pixels contribute when they are lensed by our best-fit model. This process was repeated 100 times using a randomised source plane pixelisation for each, to produce an averaged magnification profile. The total magnification is computed by calculating the flux in the source plane (for pixels above a surface brightness threshold) and the flux in the image plane due to the lensed image of the source plane pixels. Only the flux lying within the image plane mask contributes to the image plane flux and thus the total magnification.

\subsection{Intrinsic source properties}
\label{sec:source} 
\begin{table*}
	\centering
	\caption{Intrinsic source properties. The columns are total magnification, $\mu_{tot}$, dust mass computed from the single temperature optically thick SED, $M_d^{thick}$, dust mass assuming a dual temperature optically thin SED, $M_d^{thin}$, temperature of the optically thick SED, $T^{thick}$, temperatures of the optically thin SED, $T^{thin}$, the optical depth at 100 $\mu$m for the optically thick SED, $\tau_{100}$, the demagnified luminosity (computed as the integral of the optically thin SED from 3 to 1100 $\mu$m), $L_{IR}$, the $H_2$ gas mass computed using the scaling relation of \protect\cite{hughes}, $M_{gas}$, and the SFR scaled from $L_{IR}$ using the procedure given by \protect\cite{kennicut} with a Kroupa IMF. Dust masses are expressed as $\mathrm{log}(M_{\mathrm{d}}/M_{\odot})$, luminosity as $\mathrm{log}(L_{\mathrm{IR}}/L_{\odot})$, and gas masses as $\mathrm{log}(M_{\mathrm{gas}}/M_{\odot})$. The source HeLMS J235331.9+031718 appears twice, displaying the intrinsic source properties for the upper and lower value of source redshift being considered (indicated in the ID column). }
	\label{tab:sources}
	\begin{tabular}{lccccccccr} 
		\hline
		ID & $\mu_{tot}$ & $M_d^{thick}$ & $M_d^{thin}$ & $T^{thick}$ (K) & $T^{thin}$ (K) & $\tau_{100}$ & $L_{IR}$ & $M_{gas}$ & SFR ($\mathrm{M_{\odot}}$/yr)\\
		\hline
		H-ATLAS J083051.0+013225 & 6.7 $\pm$ 0.3 & 8.5 & 9.2 & 70 & 32 / 70 & 3.1 & 13.4 $\pm$ 0.1 & 11.4 $\pm$ 0.1 & 3700 $\pm$ 500 \\
		HeLMS J005159.4+062240 & 10.2 $\pm$ 0.3 & 8.2 & 8.9 & 62 & 28 / 59 & 6.9 & 12.6 $\pm$ 0.1 & 11.1 $\pm$ 0.1 & 590 $\pm$ 50 \\
		HeLMS J234051.5-041938 & 6.7 $\pm$ 0.3 & 8.3 & 9.1 & 62 & 26 / 53 & 7.8 & 13.1 $\pm$ 0.1 & 11.4 $\pm$ 0.1 & 1900 $\pm$ 300 \\
		HeLMS J232439.5-043935 & 4.9 $\pm$ 0.3 & 8.5 & 9.1 & 63 & 30 / 55 & 4.2 & 13.0 $\pm$ 0.1 & 11.3 $\pm$ 0.1 & 1500 $\pm$ 300 \\
		HeLMS J233255.4-031134 & 5.5 $\pm$ 0.3 & 8.5 & 9.1 & 60 & 28 / 51 & 4.9 & 13.1 $\pm$ 0.1 & 11.5 $\pm$ 0.1 & 1900 $\pm$ 300 \\
		HeLMS J233255.6-053426 & 9.2 $\pm$ 0.3 & 8.5 & 9.0 & 48 & 28 / 63 & 4.9 & 12.5 $\pm$ 0.1 & 11.2 $\pm$ 0.1 & 500 $\pm$ 50 \\
		HeLMS J235331.9+031718 (z=2.0) & 8.7 $\pm$ 0.3 & 8.7 & 9.0 & 37 & 25 / 31 & 3.3 & 12.1 $\pm$ 0.1 & 11.1 $\pm$ 0.1 & 180 $\pm$ 50  \\
		HeLMS J235331.9+031718 (z=3.7) & 8.7 $\pm$ 0.3 & 8.2 & 8.5 & 59 & 20 / 40 & 1.4 & 12.7 $\pm$ 0.1 & 10.8 $\pm$ 0.1 & 800 $\pm$ 50  \\
		\hline
	\end{tabular}
\end{table*}
For each lens system we have determined the intrinsic properties of the background source. To achieve this, we have demagnified the sub-mm photometry (see Table \ref{tab:photometry}) by the total source magnification factors given in Table \ref{tab:sources}, taking into account the uncertainties on our magnification values. With the source redshifts given in Table \ref{tab:redshifts}, we fitted the rest-frame photometry with two Spectral Energy Distributions (SEDs). Firstly, a single temperature optically thick SED of the form 
\begin{equation}
    S_{\nu} \propto [1 - \mathrm{exp}\left( - (\nu / \nu_0)^{\beta} \right)]B(\nu, T_d),
\end{equation}
where $S_{\nu}$ is the flux density at frequency $\nu$, $\nu_0$ is the frequency at which the optical depth equals unity, $\beta$ is the dust emissivity index, $T_d$ is the dust temperature and $B(\nu, T_d)$ denotes the Planck function. Secondly, a dual temperature optically thin SED was fitted of the form
\begin{equation}
    S_{\nu} = \nu^{\beta}[N_c B(\nu, T_c) + N_w B(\nu, T_w)],
\end{equation}
where $N$ is the weighting of the cold and warm components (denoted with subscripts) and $T$ is the dust temperature of the two components (denoted by the subscripts). This allows for the computation of an upper and lower bound in the range of possible dust masses, which were determined using the method described in \cite{dunne_2011}. In this work we have used the 873 $\mu$m ALMA flux density and computed the dust mass absorption coefficient by extrapolating the 850 $\mu$m value of $\kappa_{850} = 0.077 \mathrm{m^2 kg^{-1}}$ \citep{james} (see \cite{dunne_2000} for more details). The uncertainties in the dust mass absorption coefficient are known to be large, but a relative comparison between dust masses computed with this value is still valuable. The same value of $\kappa_{873}$ was used for all of our galaxies, which assumes that the physical properties of the dust, such as grain size and density, are constant for our sample of galaxies. The scaling relations from \cite{hughes} were used to calculate the H$_2$ gas mass.

\begin{figure*}
	\includegraphics[width=\textwidth]{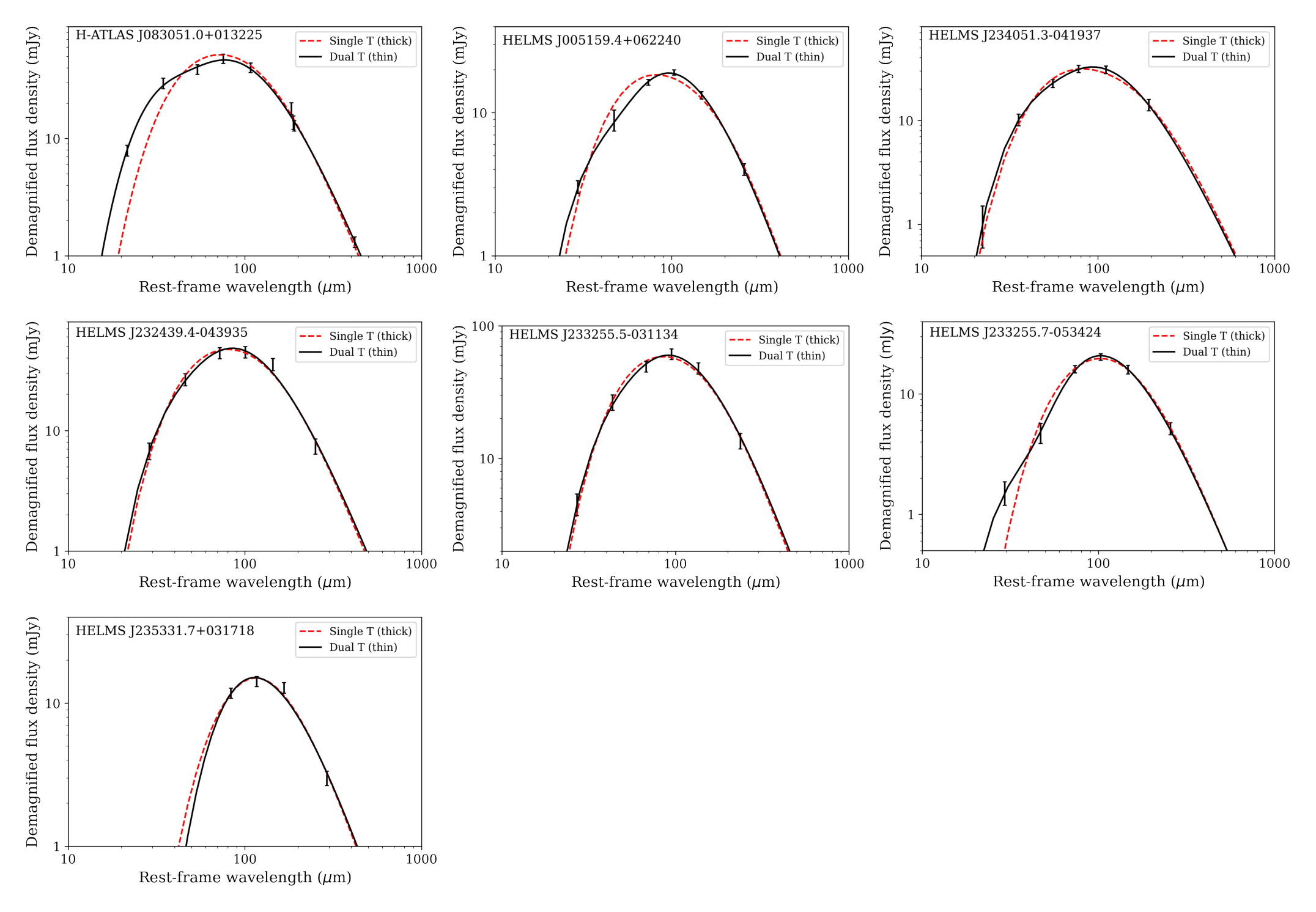}
    \caption{SEDs of the lensed background sources. The two-temperature optically thin fit (solid black line) and the single-temperature optically thick fit (dashed red line) are shown in each subplot. The measured photometry shown as the error bars in the plots have been demagnified by the appropriate lensing magnifications $\mu$, given in Table \ref{tab:sources}. The SED for HeLMS J235331.7+031718 is shown at a redshift of $z=2$.}
    \label{fig:SEDs}
\end{figure*}

During the fitting of the optically thin SED, the temperature and normalisation of both dust components were allowed to vary. In the case of the optically thick SED, the temperature, normalisation and opacity at 100 $\mu$m, $\tau_{100}$, were varied during the fit. Throughout, the emissivity index, $\beta$, was fixed to 2.0 \citep[see][]{smith}. The fitted SEDs and the demagnified source photometry can be seen in Fig \ref{fig:SEDs}. The best fit parameters for these SEDs can be found in Table \ref{tab:sources}, along with the demagnified luminosity of the sources, computed by integrating the optically thin SED from 3-1100 $\mu$m. Finally, the star formation rates of the sources are given, computed using the conversion from luminosity
\begin{equation}
    \mathrm{log(SFR) = log(L_{IR}) - 43.41}
\end{equation}
given by \cite{kennicut} which uses a Kroupa \citep{kroupa} Initial Mass Function (IMF). This conversion assumes that all of the infrared emission originates from star forming regions, which may result in biased SFR values if an AGN is significantly contributing to the luminosity of the galaxy.

\begin{figure}
	\includegraphics[width=\columnwidth]{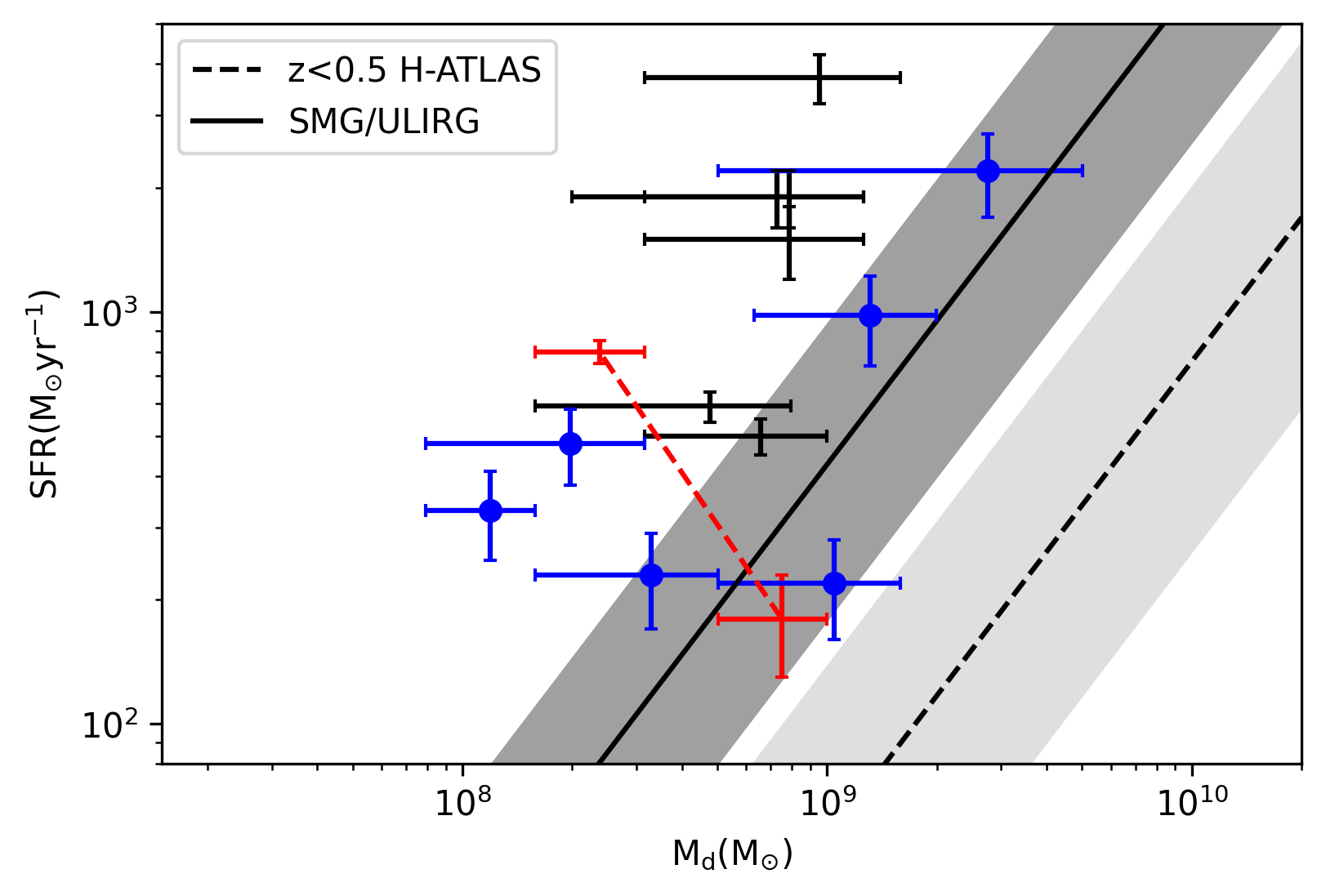}
    \caption{Star formation rate, computed with the method of \protect\cite{kennicut}, plotted against dust masses for our seven lensed sources (black pluses), with the six sources from \protect\cite{dye_2018} shown as blue pluses with a central circle. The horizontal error bars represent the range of dust masses for each source encompassed by $M_d^{thick}$ and $M_d^{thin}$, whilst the vertical error bars represent the uncertainty in SFR, plotted at the midpoint of the dust mass range. Shown in red are the values for the two extremes of redshift for HeLMS J235331.9+031718. Also shown are the empirical relations between $M_d$ and SFR determined by \protect\cite{rowlands} for high-redshift SMGs and low redshift ULIRGs (solid line with 1$\sigma$ spread indicated in dark grey) and the population of galaxies detected in H-ATLAS with $z < 0.5$ (dashed line with $1\sigma$ spread indicated in light grey).}
    \label{fig:sfr_dm}
\end{figure}

\subsection{Object notes}

\textbf{H-ATLAS J083051.0+013225} - This lens system has an almost complete $\approx 1.5$ arcsec Einstein ring with three major image components, along with a fainter central image. The central image is due to the rare line of sight configuration of the dual deflectors. Keck Adaptive Optics $K_s$ band imaging of this lensing system \citep[see][]{calanog_2014} shows two galaxies, but the lensing features are ambiguous due to the low signal to noise ratio. Upon superimposing the lensed emission from the background source as detected in our ALMA data, it is clear that both of these galaxies are interior to the Einstein ring. Long slit spectroscopy presented in \cite{bussmann_2013} provides evidence that these two galaxies are at different redshifts (0.626 and 1.002 respectively). \cite{bussmann_2013} presents the modelling of SMA data for this system, fitting two SIE models for the foreground deflectors and a Sérsic model for the background source. They infer a magnification factor $\mu = 6.9 \pm 0.6$, which is in agreement with the magnification we find with our best fit model of $\mu = 6.7 \pm 0.5$. \cite{yang} presents ALMA band 4 data, along with the band 7 data presented in this paper, and finds a significantly higher magnification factor of $\mu = 10.5^{+0.5}_{-0.6}$, using a double SIE lens model and a dual disk model for the background source.

The star formation rate of this lensed source is extremely high at $3700 \pm 500 M_{\odot}\mathrm{yr}^{-1}$, in reasonable agreement with the value reported by \cite{zhang}, who used the magnification factor from \cite{bussmann_2013} to determine the SFR. This compares to the lower values of SFR inferred by \cite{yang} who find $600 \pm 300 M_{\odot}\mathrm{yr}^{-1}$ and $900 \pm 400 M_{\odot}\mathrm{yr}^{-1}$ for each of the components in their source model. Our value of star formation rate is reduced to 2400 $M_{\odot}\mathrm{yr}^{-1}$ when using the magnification factor of 10.5 found by \cite{yang}. A possible explanation for the large discrepancy in SFR is differential magnification of the source. \cite{yang} find evidence of differential magnification for the compact and extended components of their source model, which they have taken into account in their SED modelling. In addition to this, whilst the lens model that we find is similar to that of \cite{yang}, we have allowed the power-law slope to vary rather than using SIE profiles. The higher values (>2) of slope that we find will tend to decrease the overall magnification of the source plane.

Our reconstructed source for this system shows significant disturbance, with a main component oriented North-South and a fainter component oriented East-West.

\textbf{HeLMS J005159.4+062240} - This system is a group-scale lens with a doubly-imaged background source. Optical imaging from the Sloan Digital Sky Survey (SDSS) Data Release 14 \citep{sdss} reveals two lensing galaxies within the lensed arcs. \cite{tergolina} investigated these objects, finding that one was a red and passive early-type galaxy (ETG) ($z=0.60246\pm0.00004$), whilst the other is the host of a quasar ($z=0.59945\pm0.00009$) \citep{z_helms18}. We were able to find a lens model constructed from two elliptical power-law density profiles and an external shear field that reconstructs the background source as a single component. Our reconstructed source appears elongated with a disturbed morphology.

With a magnification factor of $\mu = 10.2 \pm 0.3$, this system has the highest total magnification in our sample. Our measurement of the intrinsic source luminosity, $\mathrm{log}(L_{\mathrm{IR}}/L_{\odot}) = 12.6 \pm 0.1$, is in reasonable agreement with the far-IR luminosity given in \cite{Nayyeri_2016}, once the lens magnification has been taken into account. The peak of the SED is bounded by the SPIRE photometry and the PACS flux densities indicate the presence of a warm dust component. As such, the two temperature SED provides a better fit to the data. With a star formation rate of $\approx 590 \pm 50 M_{\odot}\mathrm{yr}^{-1}$ and a dust mass range of $10^{8.2} - 10^{8.9} M_{\odot}$ this source's SFR to dust mass ratio is typical of high-redshift SMGs and low redshift ULIRGs according to the empirical relations in \cite{rowlands}.

\textbf{HeLMS J234051.5-041938} - This system displays a nearly-complete $\approx 0.75$ arcsec Einstein ring with three image components. A singular power-law with external shear provides a good fit to the data, with the power-law index preferring relatively high values of around 2.3. The source has evidence of a disturbed morphology, displaying extended faint emission to the west and east.

The peak of the source SED is constrained by our ALMA and SPIRE photometry. The PACS flux densities do not indicate the presence of a significant warm dust component, with both the single and double temperature SEDs providing excellent fits to the data.

This source's intrinsic luminosity of $\mathrm{log}(L_{\mathrm{IR}}/L_{\odot}) = 13.1 \pm 0.1$ agrees well with the luminosity calculated by \cite{Nayyeri_2016}. A star formation rate of $\approx 1900 \pm 300 M_{\odot}\mathrm{yr}^{-1}$ and a dust mass range of $10^{8.3} - 10^{9.1} M_{\odot}$ places this source above the typical dust mass to SFR ratio for high-redshift SMGs and local ULIRGs.

\textbf{HeLMS J232439.5-043935} - This system exhibits an unusual image configuration with three distinct image components, the two western most of which show faint extended features, whilst the eastern component is more compact. Two power-law density profiles with an external shear field provide an excellent fit to the data, reconstructing a source with faint extended emission to the north.

The peak in flux density for this source occurs close to the 250 $\mu$m SPIRE measurement, with the constraints coming from the PACS measurements. The SED does not show clear evidence of a significant second temperature component, with both SED models providing good fits to the measured flux densities.

With an intrinsic luminosity of $\mathrm{log}(L_{\mathrm{IR}}/L_{\odot}) = 13.0 \pm 0.1$, our estimate is consistent within our uncertainties of the value given by \cite{Nayyeri_2016}. A star formation rate of $\approx 1500 \pm 300 M_{\odot}\mathrm{yr}^{-1}$ and a dust mass range of $10^{8.5} - 10^{9.1} M_{\odot}$ means that this source has a higher than typical ratio of SFR to dust mass for high-redshift SMGs and local ULIRGs.

\textbf{HeLMS J233255.4-031134} - A doubly-imaged source with some faint extended emission emanating from the southern image, that is well fit by a single power-law density profile and an external shear field. The reconstructed source is relatively compact and featureless.

The peak of the SED is constrained by the SPIRE photometry, with the PACS flux densities not showing any major warm dust component. The SED is well described by an optically thick model.

The intrinsic source luminosity we have obtained of $\mathrm{log}(L_{\mathrm{IR}}/L_{\odot}) = 13.1 \pm 0.1$ is in agreement with \cite{Nayyeri_2016}. The SFR of $\approx 2000 \pm 300 M_{\odot}\mathrm{yr}^{-1}$ for this source agrees with the value quoted by \cite{zhang} within our stated uncertainties. A dust mass range of $10^{8.5} - 10^{9.1} M_{\odot}$ means that this galaxy also lies above the mean SFR to dust mass ratio as given by \cite{rowlands}.

\textbf{HeLMS J233255.6-053426} - This quadruple image system is well described by a power-law density profile embedded in an external shear field. The source exhibits a relatively featureless compact morphology.
The peak of the source SED is well constrained by the ALMA and SPIRE photometry. The PACS photometry indicates the presence of a warmer dust component, as shown by the relatively high 100 $\mu$m PACS flux density measurement, and the significantly better fit of the two temperature SED.
Our measurement of the intrinsic source luminosity, $\mathrm{log}(L_{\mathrm{IR}}/L_{\odot}) = 12.5 \pm 0.1$, is in agreement with the far-IR luminosity given in \cite{Nayyeri_2016} using our magnification factor of 9.2 to de-magnify the quoted value. With a star formation rate of $\approx 500 \pm 50 M_{\odot}\mathrm{yr}^{-1}$ and a dust mass range of $10^{8.5} - 10^{9.0} M_{\odot}$ this source's SFR to dust mass ratio is consistent with typical high-redshift SMGs and low redshift ULIRGs as indicated by Fig \ref{fig:sfr_dm}.

\textbf{HeLMS J235331.9+031718} - This double image system has an extremely small image separation, with an Einstein radius of $\approx 0.1$ arcsec. This system is described by a single power-law density profile and an external shear field. The inferred slope of the power-law density profile is relatively low at $\alpha = 1.64 \pm 0.04$, contributing to the high magnification of this source, which also lies near a lensing caustic cusp. The source itself appears to be mostly compact with an extended feature to the south east, which is readily visible in the observed lensed image.

There is no redshift measurement for this source and so we have opted to use the range of redshifts ($z \sim 2 - 3.7$) present in our sample to calculate redshift dependent quantities. There are also no PACS flux density measurements for this source, and so we rely on the SPIRE and ALMA measurements to constrain the SED. The peak of the SED appears to lie within the SPIRE wavelengths, and fitting the optically thick SED for the range of redshifts considered gives a temperature range of $37 -- 59$ K. Without the PACS measurements and with the extra free parameters of the optically thin SED, it is not possible to meaningfully infer the presence of a warmer dust component.

\section{Discussion}
\label{sec:discussion}

\begin{figure}
	\includegraphics[width=\columnwidth]{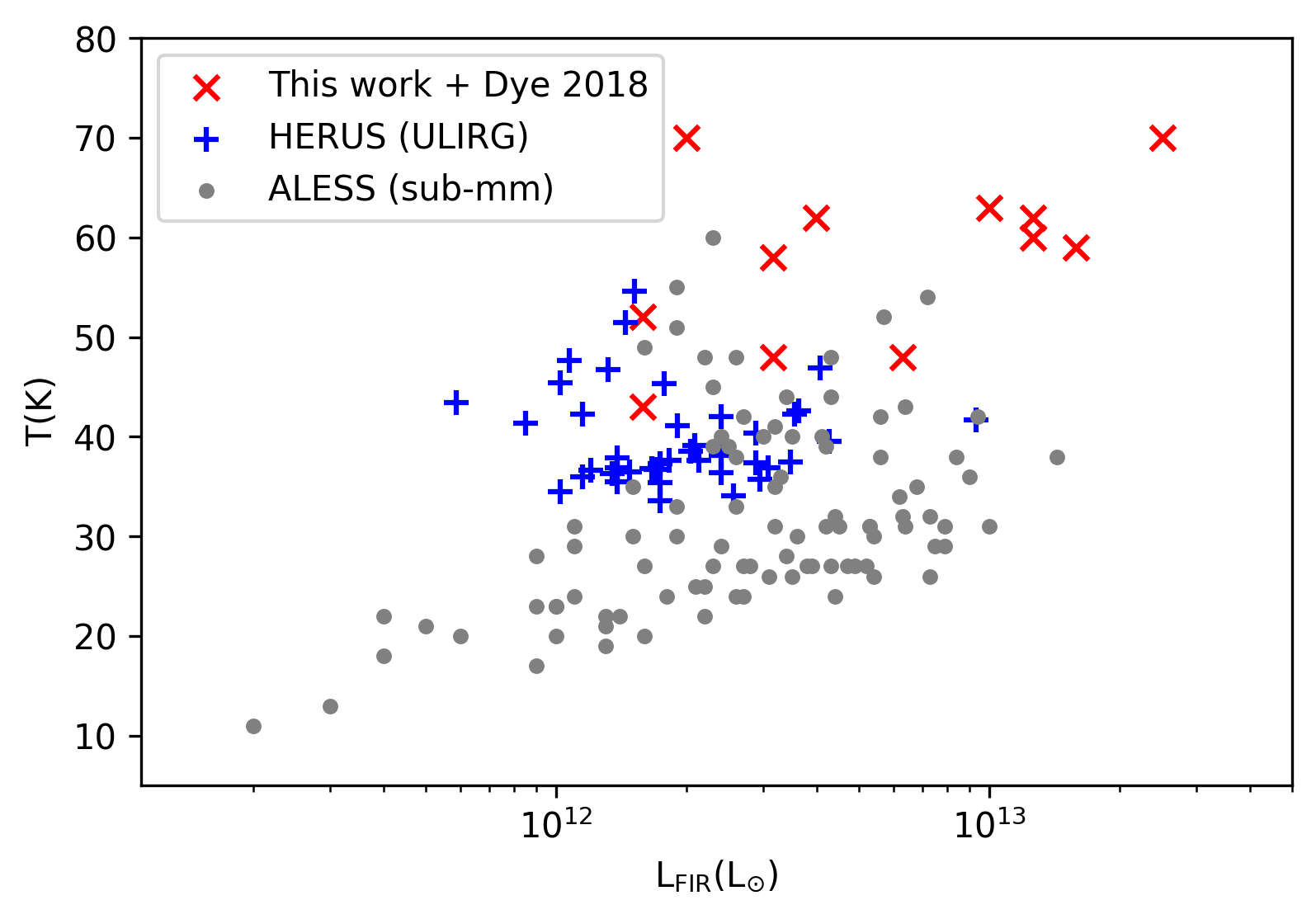}
    \caption{Temperature distribution as a function of far-IR luminosity. The red crosses show the luminosities taken from this work (excluding HeLMS J235331.9+031718) and \protect \cite{dye_2018}. The spectroscopic redshift range of this combined dataset is approximately $z \sim 1 - 4$. The blue pluses show the data taken from the HERUS Survey \protect \citep{herus} with a spectroscopic redshift range of approximately $z \sim 0.02 - 0.26$. The grey dots come from the ALESS survey \protect \citep{swinbank} with a photometric redshift range of $z \sim 0.5 - 6.5$.}
    \label{fig:L_T}
\end{figure}

Combining our sample with that of \cite{dye_2018}, who carried out a similar analysis on a set of 6 strongly lensed sub-mm galaxies observed as part of the same ALMA programme, we can start to make more significant comparisons between our results and those found by other surveys. The HERschel ULIRG Survey \citep[HERUS;][]{herus} is a sample of 43 local ULRIGS, selected at 60 $\mu$m by the Infrared Astronomical Satellite (IRAS). The ALMA LESS survey is a set of cycle 0 and cycle 1 ALMA observations of the sub-mm sources detected in the LABOCA ECDFS Submm Survey (LESS). Comparing the relationship between the far-IR luminosity and the effective dust temperature (see Fig. \ref{fig:L_T}) from the optically thick model between our sample, the ULIRG population from HERUS and the sub-mm sources in ALESS, we can see that our sources tend to possess both more extreme luminosities and higher dust temperatures. The median luminosity of the of the ALESS sample is $\mathrm{L_{IR} = (3.0 \pm 0.3) \times 10^{12} L_{\odot}}$, compared with the median luminosity of the HERUS sample of $\mathrm{L_{IR} = (1.7 \pm 0.3) \times 10^{12} L_{\odot}}$, and finally, that of our sample; $\mathrm{L_{IR} = (5.1 \pm 2.5) \times 10^{12} L_{\odot}}$. Given our small sample size and large scatter on this measurement, overall, our luminosities are consistent with both the HERUS and ALESS samples. The median effective temperature of the ALESS sample is $\mathrm{T = 31 \pm 1}$K, which is consistent at the 2-$\sigma$ level with that of the HERUS sample $\mathrm{T = 38 \pm 3}$K. The median dust temperature of our sample is significantly higher at $T = 59 \pm 3$K. The high dust temperatures can be interpreted as a product of the extreme SFRs present in our sample, though it is important to bear in mind that the significance of these results are likely explained by the combination selection effects due to our observing of the brightest galaxies detected by Herschel and the sample of ALESS galaxies being selected at 870$\mu m$.

Fig \ref{fig:sfr_dm} shows the degree to which our sources exceed this ratio by plotting the SFR determined using the scaling relation of the far-IR luminosity given by \cite{kennicut} against the dust mass. Given that our SFRs are derived from scaled far-IR luminosity, we can conclude that our sources have higher than expected luminosities for the amount of gas available for star formation. An obvious interpretation of this fact would be that a large fraction of the luminosity is due to an active galactic nucleus, but without additional imaging we cannot confirm this. It is also important to note that this excess could be at least in part explained by selection bias, as we have chosen the brightest Herschel sources to follow-up on.

Converting the rest frame 850 $\mu$m flux density of our sources to H$_2$ gas mass (see Table \ref{tab:sources}), using the scaling relation given by \cite{hughes}, we find that our sources all lie on or above the mean relationship between SFR and H$_2$ gas mass as determined by \cite{scoville}. An interpretation of this is that our sources possess a higher star formation efficiency (SFE), provided that dust is an accurate tracer for molecular gas. Fitting a line parallel to the SMG/ULIRG relation shown in Fig. \ref{fig:sfr_dm}, treating the range of dust masses as the 1-$\sigma$ error, we find an increase in SFE by a factor of 6 when compared with the value implied by the SMG/ULIRG relation from \cite{rowlands}. This factor increases to 50 when compared to the $z < 0.5$ H-ATLAS galaxies.

An unknown fraction of our sources could have their SFR over estimated due to significant contamination in the IR by strongly obscured AGN. In order to quantify this effect, additional observational evidence would need to be considered, e.g estimating the stellar mass from broadband SED fitting, which would require careful lens light subtraction or additional follow-up observations in the X-ray hard band to reveal the obscured nucleus.

\section{Conclusions}
\label{sec:conclusions}

We have modelled seven ALMA observations of strongly lensed sub-millimetre galaxies. Four of these systems are galaxy-galaxy scale strong lenses, which are well described by a single power-law mass profile, whilst the remaining three are group-scale lenses, which have been successfully fitted with two power-law mass profiles. Where we found it improved the fit, an external shear term was also included in the lens model. In this work we have opted to model the visibility data directly rather than to work with CLEANed image plane data. Whilst the uv-plane method is more computationally expensive, it does not suffer from the image pixel covariances introduced due to incomplete sampling of the uv-plane. However, \cite{dye_2018} showed that with sufficient sampling of the uv space, both the image-plane and direct visibility modelling approaches produce very similar results. We have fitted smooth power-law density profiles (and in some cases two) for each of the lensing systems, and found that most of the lenses are close to isothermal. This result is expected in massive ETGs due to the combination of an inner Sérsic profile representing the baryonic component and an outer NFW profile representing the dark matter component \citep{lapi_profile}. In some instances, there are significant deviations from an isothermal power-law slope, which may be due to degeneracies between parameters in our model or reflect the true nature of the lens, but we leave a more thorough explanation of this to further work.

By obtaining the total magnification factors from our models, we have demagnified the available sub-millimetre source photometry. Fitting rest frame SEDs to this data allowed us to determine the dust temperature, dust mass, intrinsic luminosity and star formation rates of our lensed sources. To estimate a range of possible dust masses for these sources, we fitted the photometry with both single temperature optically thick and dual temperature optically thin SEDs. Using the midpoint of this range to calculate the SFR to dust mass ratio, we find that all seven of our sources lie above the mean ratio for the SMG/ULIRG population as described in \cite{rowlands}.

Galaxy morphology is strongly correlated with star formation history \citep{larson, strateva, lee}. At $z \sim 3$ massive galaxies are mostly star forming disks, with the SFR peaking at $1 < z < 2$ \citep{madau}, and then dropping with the rapid growth of the fraction of massive quiescent galaxies. This period of galaxy evolution is extremely dramatic, and many different mechanisms have been proposed to explain the build up of of ETGs we see today. Galaxy mergers are one such mechanism as they are effective at disturbing the morphology and building a central bulge in a galaxy \citep{hopkins, snyder}. They have also been shown to trigger starbursts and AGN, which can lead to strong supernovae and/or AGN winds contributing to the quenching of a galaxy \citep{bekki}. Our combined sample of strongly lensed sub-mm galaxies contains a majority of sources that display a disturbed morphology. 9 of the 13 galaxies in our sample are visually classified as being disturbed, with at least two of them having evidence of being mergers (H-ATLAS J142935.3-002836, H-ATLAS J083051.0+01322). Other observations of sub-mm galaxies have found similarly high fractions of disturbed morphologies, such as \cite{chapman}, who used high-resolution optical and radio imaging of 12 sub-mm galaxies to study their spatially extended star formation activity. It has been suggested that high density molecular gas is more commonly found in galaxy mergers than quiescent systems and that this can be used to predict the star formation mode of a galaxy \citep{papadopoulos}. We are not able to conclusively say what fraction of our sample's disturbed morphologies are a result of mergers, but the source with the most extreme ratio of SFR and gas mass (H-ATLAS J083051.0+01322) does display a significantly disturbed morphology and is identified as being a merger by \cite{yang}.

\section*{Acknowledgements}
JM acknowledges the support of the UK Science and Technology Facilities Council (STFC). CF acknowledges financial support from CNPq (processes 433615/2018-
4 e 314672/2020-6). JGN acknowledges the PGC 2018 project PGC2018-101948-B-I00 (MICINN/FEDER). MG acknowledges the support of the UK STFC. AL is partly supported by the PRIN MIUR 2017 prot.20173ML3WW 002 ‘Opening the ALMA window on the cosmic evolution of gas, stars, and massive black holes’, and by the EU H2020-MSCA-ITN-2019 project 860744 ‘BiD4BEST: Big Data applications for Black hole Evolution STudies’. JLW acknowledges support from an STFC Ernest Rutherford Fellowship (ST/P004784/1 and ST/P004784/2).

This paper makes use of the following ALMA data: ADS/JAO.ALMA\#2013.1.00358.S. ALMA is a partnership of ESO (representing its member states), NSF (USA) and NINS (Japan), together with NRC (Canada) and NSC and ASIAA (Taiwan) and KASI (Republic of Korea), in cooperation with the Republic of Chile. The Joint ALMA Observatory is operated by ESO, AUI/NRAO and NAOJ. 
\section*{Data Availability}
The data underlying this article were accessed from the ALMA Science Archive, and are contained within the dataset  ADS/JAO.ALMA\#2013.1.00358.S. The derived data generated in this research will be shared on reasonable request to the corresponding author.


\bibliographystyle{mnras}
\bibliography{example} 





\bsp	
\label{lastpage}
\end{document}